\documentclass[prb,showpacs,floatfix,twocolumn]{revtex4}\usepackage{graphicx}

\newcommand{\srolro}{(SrRuO$_3$)$_{1-x}$(LaRhO$_3$)$_x$}
\newcommand{\sro}{SrRuO$_3$}
\newcommand{\lro}{LaRhO$_3$}
\newcommand{\cro}{CaRuO$_3$}
\newcommand{\srocro}{Sr$_{1-x}$Ca$_x$RuO$_3$}

\begin{document}

\title{Electrical and magnetic properties of the complete solid solution series between \sro\/ and \lro: Filling $t_{2g}$ \textit{versus} tilting} 

\author{Phillip T. Barton} \email{pbarton@mrl.ucsb.edu}
\author{Ram Seshadri} \email{seshadri@mrl.ucsb.edu}
\affiliation{Materials Department and Materials Research Laboratory\\
University of California, Santa Barbara, CA, 93106, USA}

\author{Matthew J. Rosseinsky} \email{M.J.Rosseinsky@liverpool.ac.uk}
\affiliation{Department of Chemistry, The University of Liverpool \\
L69 7ZD, UK}

\date{\today}

\begin{abstract}

A complete solid solution series between the $t_{2g}^4$ perovskite ferromagnet \sro\/ and the diamagnetic $t_{2g}^6$ perovskite \lro\/ has been prepared. The evolution with composition $x$ in \srolro\/ of the crystal structure and electrical and magnetic properties has been studied and is reported here. As $x$ increases, the octahedral tilt angle gradually increases, along with the psuedocubic lattice parameter and unit cell volume. Electrical resistivity measurements reveal a compositionally driven metal to insulator transition between $x$ = 0.1 and 0.2. Ferromagnetic ordering gives over to glassy magnetism for $x \geq 0.3$ and no magnetic ordering is found above 2\,K for $x > 0.5$. $M_{\rm{sat}}$ and $\Theta_{CW}$ decrease with increasing $x$ and remain constant after $x$ = 0.5.  The magnetism appears poised between localized and itinerant behavior, and becomes more localized with increasing $x$ as evidenced by the evolution of the Rhodes-Wohlfarth ratio. $\mu_{\rm{eff}}$ \textit{per} Ru is equal to the quenched spin-only $S$ value across the entire solid solution. Comparisons with \srocro\/ reinforce the important role of structural distortions in determining magnetic ground state. It is suggested that electrical transport and magnetic properties are not strongly coupled in this system.


\end{abstract}

\maketitle 

\section{Introduction} 

\sro\/ is a 4$d$ ($t_{2g}^4 \, e_g^0$) transition metal oxide crystallizing in the orthorhombic ABO$_3$ perovskite structure.\cite{Randall_JACS1959,Callaghan_IC1966,Longo_JAP1968} Unusually for a 4$d$ transition metal oxide, \sro\/ exhibits ferromagnetic ordering below its Curie temperature $T_c$ of 160\,K. This is in contrast to other 4$d$ perovskites such as (Ca,Sr,Ba)MoO$_3$ whose metal $d$ -- O $p$ conduction bands are too disperse to stabilize magnetic ordering. Closely related CaRuO$_3$ is, in contrast to SrRuO$_3$, a paramagnetic metal while BaRuO$_3$ orders ferromagnetically at a lower $T_c$ of 60\,K.\cite{Longo_JAP1968,Jin_PNAS2008} Recent renewed interest in SrRuO$_3$ has origins in the unusual negative spin polarization, as determined from tunneling measurements,\cite{Worledge_PRL2000} that enables multilayer devices with inverted magnetoresistance behavior by combination with materials exhibiting the more usual positive spin polarization.\cite{Takahashi_PRB2003} The nature of magnetism in SrRuO$_3$, localized versus itinerant, continues to be examined using chemical substitution as a probe.\cite{Jin_PNAS2008} Additionally, as SrRuO$_3$ is a commonly used electrode material in heteroepitaxial perovskite architectures, recent attention has been paid to the thickness-dependence of properties.\cite{Toyota_APL2005,Chang_PRL2009}

Jones \textit{et al.}\cite{Jones_ACC1989} performed room temperature neutron diffraction studies studies to determine the structure of \sro\/ while Bushemeleva \textit{et al.}\cite{Bushmeleva_JMMM2006} performed low temperature experiments and obtained a magnetic moment per Ru atom of 1.63$\pm$0.06\,$\mu_B$ at 10\,K. They also explained a previously observed Invar effect\cite{Kiyama_PRB1996} of the lattice parameters $a$ and $b$ below below $T_c$ as deriving from the freezing of RuO$_6$ octahedra tilting and rotation. Additionally, a very slight Jahn-Teller distortion, 40 times smaller than in LaMnO$_3$, was noted.\cite{Bushmeleva_JMMM2006}

The reported Invar effect in SrRuO$_3$ is indicative of itinerancy as it also occurs in 3$d$ transition metal itinerant ferromagnets such as Fe-Ni alloys.\cite{Kiyama_PRB1996} Values from heat capacity, $\Delta C$ and $\Delta S$,\cite{Allen_PRB1996,Castelpoggi_SSC1997} and the NMR relaxation rate 1/$T_1$\cite{Yoshimura_PRL1999} are all lower than the values expected from the localized spin model. Many electronic structure calculations predict band ferromagnetism with a reduced non-integer moment and this is consistent with the precise moment value extracted from low temperature neutron scattering experiments. Additionally, the linearity of the Arrot plot for small $H/M$ predicted for itinerant magnets is observed in \sro.\cite{Kiyama_JPSJ1999}

Extensive computational investigations of the magnetism and electronic structure of \sro\/ and its behavior in solid solutions have been carried out.\cite{Mazin_PRB1997,Santi_JPCM1997,Rondinelli_PRB2008} Despite its itinerant nature, Mazin and Singh emphasize the importance of structural distortion in \sro\/ which changes the Ru--O--Ru bond angle and affects ferromagnetic coupling.\cite{Mazin_PRB1997} A large peak is seen in the density of electronic states at the Fermi level which supports magnetic ordering \textit{via} the Stoner criterion.\cite{Stoner_PRSL1938} In the more distorted \cro\/ the density of states is not as strongly peaked, and it therefore displays no magnetic order but is believed to be on the verge of a ferromagnetic instability. Other LSDA studies of \sro\/ and \cro\/ by Santi \textit{et al.} show both to be ferromagnetic.\cite{Santi_JPCM1997} Rondinelli \textit{et al.} find that their computed electronic structure of \sro\/ agrees better with experimental spectroscopic data when moderate electron correlations are included through a 0.6\,eV on-site Hubbard term.\cite{Rondinelli_PRB2008} Maiti and Singh fit photoemission data and find a small $U/W$ of 0.2 for both \cro\/ and \sro\/ which similarly suggests these compounds are not significantly correlated.\cite{Maiti_PRB2005} Calculations comparing ideal cubic structures and real distorted structures reveal the important role of A--O covalency and Ru--O--Ru bond angle.\cite{Maiti_PRB2006}

In recent work on \sro\/ solid solutions, Mamchik \textit{et al.} investigated substitution by antiferromagnetic LaFeO$_3$\cite{Mamchik_PRB2004_A} and LaCoO$_3$.\cite{Mamchik_PRB2004_B} In both cases, a spin glass forms upon substitution and a gradual metal-insulator transition is suggested to occur by Anderson localization. Additionally, large switchable local moments are formed around the substituted B-site (Fe or Co) due to the spin polarization of the itinerant electrons in \sro. An analogy with 3$d$ transition metal impurities in Pd has been suggested and correspondingly, and large negative magnetoresistance is reported. Pb substitution has been attempted on both the A and B sites with mixed results. Cao \textit{et al.} perform substitution, ostensibly on the B site, and report an increase in $T_c$ to 210\,K.\cite{Cao_PRB1996} Cheng \textit{et al.} on the other hand, replace Sr with Pb on the A site and observe a reduction in $T_c$ with no magnetic ordering for $x$ = 0.6 or higher.\cite{Cheng_PRB2010} The effect of structural distortion on the magnetism of \sro\/ has been of great interest for the past 50 years. Most simply accomplished by isovalent substitution on the A site, replacement of Sr by both Ca and Ba has been investigated. \srocro\/ has been the subject of a sizable number of experimental on both single crystals and polycrystalline samples, and through computational studies. However, the results are somewhat varied.\cite{Kanbayasi_JPSJ1978,Fukunaga_JPSJ1994,Cao_PRB1997,Kiyama_JPSJ1998,Yoshimura_PRL1999,He_PRB2000,Jin_PNAS2008,Longo_JAP1968,Kiyama_PRB1996,Mukuda_PRB1999,Cao_SSC2008} Specifically, there are reports of glassy magnetic ordering persisting to $x$ = 0.95,\cite{Cao_PRB1997} while most studies seem to agree on a value of $x$ = 0.7, beyond which no magnetic ordering is observed. Ba substitution, on the other hand, is harder to perform as high pressure is required to stabilize BaRuO$_3$ in the perovskite phase. Recently, the entire solid solution series ARuO$_3$ (A = Ca, Sr, Ba) was studied by Jin \textit{et al.}\cite{Jin_PNAS2008} These authors report that the Curie temperature decreases with substitution of either Ca or Ba, and they attribute this to many competing effects, including changes in octahedral tilting and rotation, Jahn-Teller distortions, and covalence. With Ba substitution, $T_c$ suppression is explained by band broadening. In the case of Ca substitution, these authors report the formation of a Griffiths phases, characterized by some signature  in the paramagnetic susceptibility at the ordering temperature of the parent phase, $T_G$, due to local clusters of the ferromagnet persisting in the dilute system.\cite{Griffiths_PRL1969,Bray_PRL1987} Kiyama \textit{et al.} have also suggested that Sr$^{2+}$ clustering occurs in \srocro\/ but do not report signals in the susceptibility at $T_G$.\cite{Kanbayasi_JPSJ1978} Additionally, they found that long range magnetic ordering persists through $x$ = 0.7.\cite{Kiyama_JPSJ1999}

In this contribution, we investigate \srolro\/, a solid solution between \sro\/ and the perovskite \lro\/ which has low-spin, diamagnetic $d^6$ Rh$^{3+}$.\cite{Wold_JACS1957,Nakamura_JSSC1993,Mary_JSSC1994} Aliovalent substitution on both the A and B sites of the ABO$_3$ perovskite compensates Ru$^{4+}$ being replaced by Rh$^{3+}$ with the concomitant substitution of Sr$^{2+}$ by La$^{3+}$. The principle electronic effect anticipated by such substitution, at least in a band picture, would be the gradual filling of $t_{2g}$ levels on the B site, starting with $t_{2g}^4$ SrRuO$_3$ at $x$ = 0, and ending with $t_{2g}^6$ SrRuO$_3$ at $x$ = 1.  While \sro\/ has been substituted by many different ions as described above, this is the first time substitution by a diamagnetic semiconductor has been attempted. The only example of Rh-substitution is from Cao \textit{et al.}\cite{Cao_JAP1997} who have substituted Rh (ostensibly Rh$^{4+}$) for Ru in CaRuO$_3$ and found that it stabilized magnetic ordering. The results presented here shed important light on the nature of magnetism in \sro\/. Itinerant behavior becomes more localized with substitution by a semiconductor as evidenced by the evolution of the Rhodes-Wohlfarth ratio\cite{Rhodes_PRSL1963} with $x$ in \srolro\/. Additionally, despite the occurrence of a compositionally driven metal-insulator transition, ferromagnetism persists, with behavior turning glassy as $x$ increases. Curie-Weiss analysis reveals that $\Theta_{CW}$ decreases with \lro\/ substitution and is equal to $T_c$ for low valuies of $x$, while $\mu_{\rm{eff}}$ \textit{per} Ru$^{4+}$ is equal to the spin-only $S$ value across the solid solution. Interestingly, we find that the ferromagnetism of \sro\/ is disrupted only slightly more quickly by \lro\/ substitution than it is with isovalent replacement of Sr by Ca. This suggests that increased octahedral tilting is almost as detrimental to ferromagnetism as is filling $t_{2g}$, and reinforces prior studies that emphasize the significance of structural distortion.

\section{Methods}

Polycrystalline \srolro\/ pellets were prepared using solid state reactions at high temperatures. Stoichiometric amounts of SrCO$_3$, La$_2$O$_3$, RuO$_2$, and Rh$_2$O$_3$ were ground with an agate mortar and pestle, pressed at 100\,MPa, and fired in air, first at 1000$^{\circ}$C for 24\,h and then between 1200$^{\circ}$C and 1400$^{\circ}$ for 96\,h with intermediate grindings in accordance with previous preparations of \sro\cite{Randall_JACS1959} and \lro\cite{Wold_JACS1957}. The pellets were placed on beds of powders of the same composition to avoid contamination with crucible constituents. The phase purity of all samples was confirmed by laboratory x-ray diffraction on a Philips X'Pert diffractometer with Cu-$K_\alpha$ radiation. Select samples were also examined by high resolution synchrotron powder x-ray diffraction at the 11-BM beamline at the Advanced Photon Source, Argonne National Laboratory. Rietveld\cite{Rietveld_JAC1969} refinement was performed using the XND Rietveld code.\cite{berar_xnd_1998} Crystal structures were visualized using VESTA.\cite{momma_vesta_2008} Electrical resistivity measurements were carried out using the 4-point probe method on sintered pellets with silver epoxy electrical contacts using a Quantum Design PPMS, and separately using Keithley current sources and meters and a closed-cycle He refrigerator. Magnetic properties were measured using a Quantum Design MPMS 5XL SQUID magnetometer and the ACMS option in a PPMS. 

\section{Results and Discussion}

\subsection{Structure}

\begin{figure}
\centering\includegraphics[width=3in]{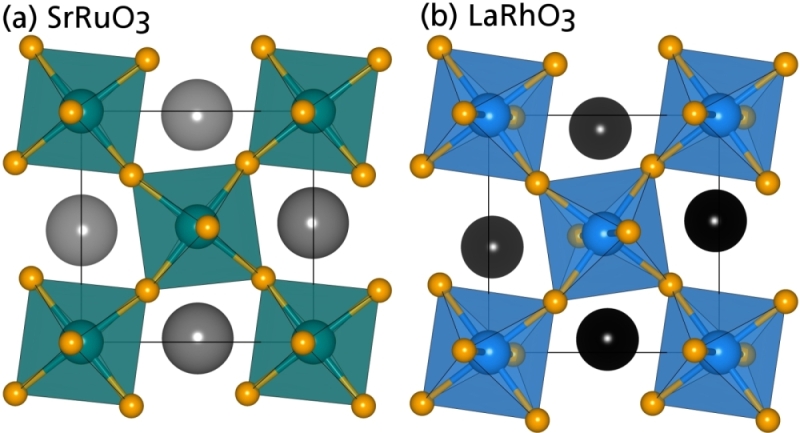}\\ \caption{(Color online) Orthorhombic perovskite crystal structures of (a) \sro\/ and (b) \lro\/ viewed down the long $b$ axis. The structures were determined by Rietveld refinement of powder XRD data. \lro\/ has significantly more tilting than \sro, shown by the displacement of the apical O, while the rotation angles are similar. The sphere colors correspond to: grey, Sr; black, La; green Ru, blue Rh, and orange O. 
\label{fig:structures}}
\end{figure}

\begin{figure}
\centering\includegraphics[width=3in]{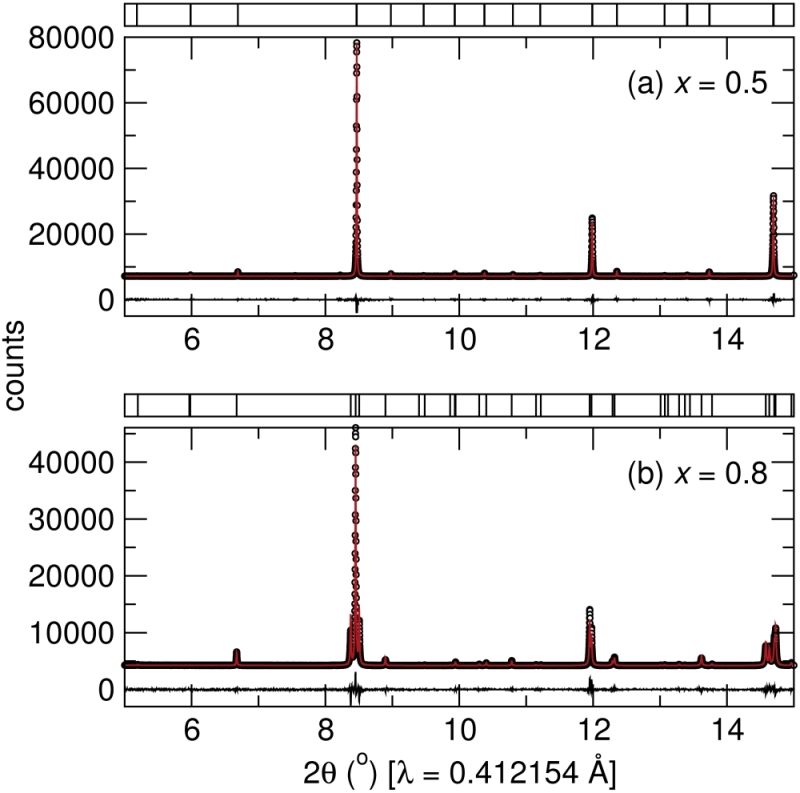}\\
\caption{(Color online) 
High resolution synchrotron powder x-ray diffraction data and Rietveld refinement for (a) $x$ = 0.5 and (b) $x$ = 0.8 in \srolro. Data (circles), the Rietveld fit (red lines, $R_{\rm{Bragg}}$ $<$ 7\,\% for all samples), and difference between data and fit are displayed. Vertical lines at the top of the panels indicate expected peak positions. The top and bottom panels show data for $x$ = 0.5 and $x$ = 0.8 respectively.
\label{fig:sync}}
\end{figure}

\begin{figure}
\centering\includegraphics[width=3in]{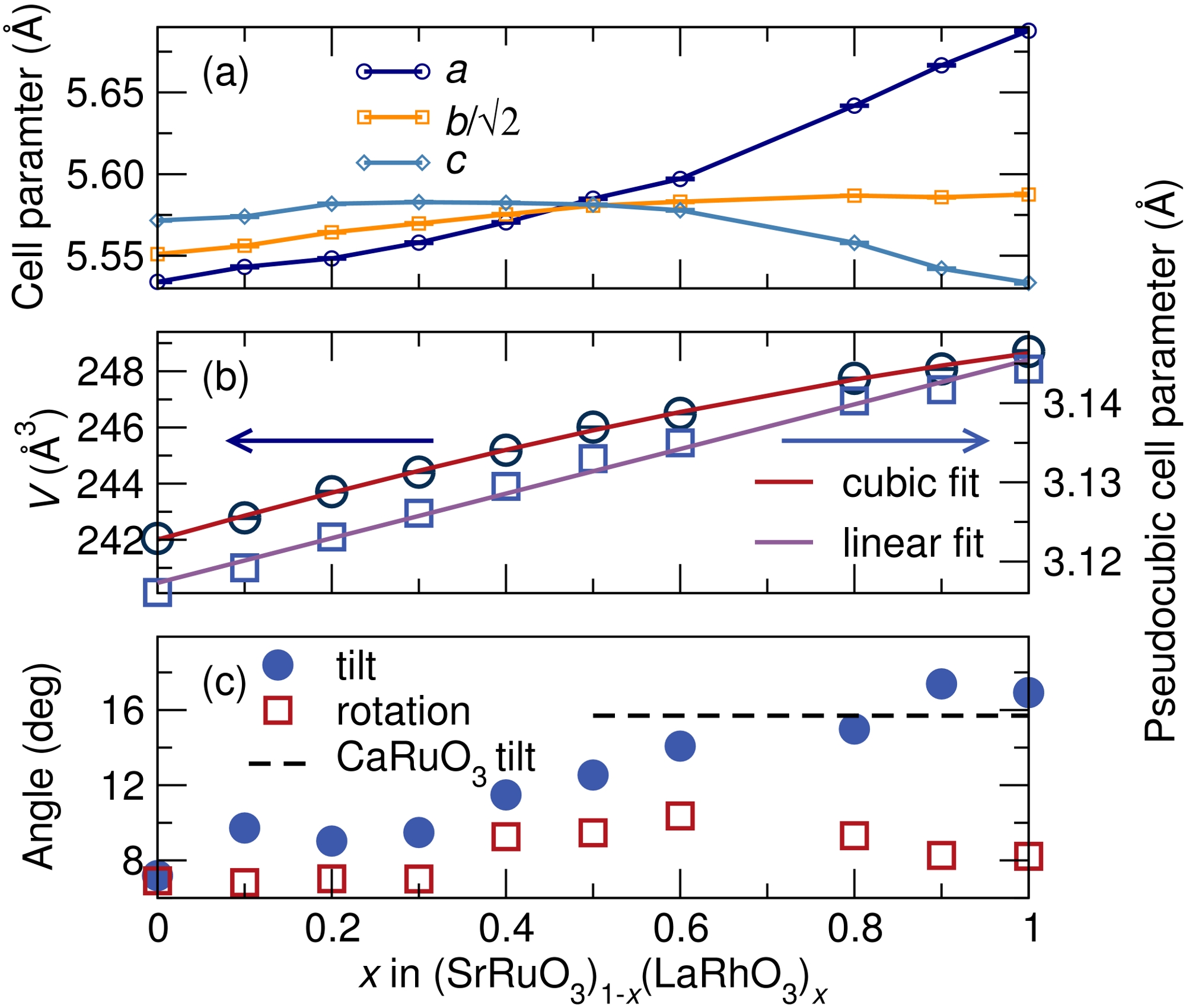}\\
\caption{(Color online) 
(a) Unit cell parameters as a function of composition. (b) Unit cell volume (circles) and psuedocubic cell parameter (squares) as a function of composition. The lines are cubic and linear fits to the data, demonstrating that the V\'egard law is followed. (c) Perovskite tilt $\phi$ and rotation $\theta$ angles as a function of composition. BO$_6$ octahedra are increasingly tilted with \lro\ substitution, while the rotation angle remains mostly constant.
\label{fig:struct_params}}
\end{figure}

\begin{table*}
\caption{Unit cell parameters and cell volume for \srolro\, obtained from Rietveld refinement of powder x-ray diffraction data in space group 
$Pnma$ (No. 62). \label{tab:struc}}
\centering
\begin{tabular}{lccccccccccccc}
\hline
\hline
$x$ &0.0 &0.1 &0.2 &0.3 &0.4 &0.5 &0.6 &0.8 &0.9 &1.0 \\
\hline
$a$ (\AA) &5.534(0) &5.543(2) &5.548(4) &5.558(0) &5.570(4) &5.584(9) &5.597(0) &5.641(9) &5.666(6) &5.687(8)\ \\
$b$ (\AA) &7.850(3) &7.857(5) &7.869(2) &7.876(8) &7.884(7) &7.892(2) &7.895(7) &7.900(9) &7.899(6) &7.902(0)\ \\
$c$ (\AA) &5.571(6) &5.574(0) &5.581(8) &5.582(8) &5.582(4) &5.581(4) &5.577(8) &5.557(9) &5.542(1) &5.533(2)\ \\
$V$ (\AA$^3$) &242.0(5) &242.7(8) &243.7(1) &244.4(1) &245.1(9) &246.0(1) &246.5(0) &247.7(5) &248.0(8) &248.7(0)\ \\
\hline
\hline
\end{tabular}
\end{table*}

The crystal structures of \sro\/ and \lro\/ as determined by Rietveld refinement of powder XRD data are depicted in FIG.\,\ref{fig:structures}. Both end-member compounds crystallize in the orthorhombic perovskite crystal structure, space group $Pnma$ (No. 62), with the latter showing a greater degree of octahedral tilting. Powder XRD shows the single phase nature of the entire solid solution as all observable peaks are expected from the structure. Upon \lro\/ substitution, many of the peaks display enhanced splitting due to increased orthorhombic distortion. Figure\,\ref{fig:sync} shows high resolution synchrotron powder x-ray diffraction data and Rietveld refinement for the $x$ = 0.5 and 0.8 samples. The high quality data further confirm the single phase nature of the materials and refined La/Sr ratios agree well with stoichiometry. The refined cell parameters and unit cell volume are presented in Table\,\ref{tab:struc}. The composition dependence of structural parameters is shown in FIG.\,\ref{fig:struct_params}. The individual cell parameters do not follow the V\'egard law due to the effect of octahedral rotations and tilts. Instead, the unit cell volume and psuedocubic cell parameter follow the V\'egard law as evidenced by the cubic and linear fits to the data. The observation that the V\'egard law is obeyed across the solid solution strongly suggests that there is no change in the oxidation states of the transition metal ions Ru$^{4+}$ or Rh$^{3+}$ as $x$ changes in the solid solution. The bottom panel, FIG.\,\ref{fig:struct_params}(c), displays the average octahedral tilt angle as a function of composition. It is seen that octahedral tilting increases with \lro\/ substitution as expected based on the larger A-cation charge and therefore the smaller tolerance factor:\cite{Goldschmidt_1926} $t = (r_A+r_O)/(\sqrt{2}(r_B+r_O))$ is 0.994 for \sro\/ and $t$ = 0.945 for \lro\/ using Shannon-Prewitt\cite{Shannon_ACB1969} effective ionic radii. The perovskite tilt and rotation systems are described by Glazer, with $Pnma$ belonging to the $a^-b^+a^-$ tilt system.\cite{Glazer_Acta1972} For comparison to \srolro\/, the tilt angle of CaRuO$_3$, which is nearly the same magnitude as in \lro\/, is displayed as a horizontal dashed line. Since Ca substitution on the Sr site of SrRuO$_3$ significantly influences properties, the analogous tilting in the \sro--\lro\/ solid solution is expected to have a similar effect and be an important ingredient to understanding physical properties.

\subsection{Electrical transport}

\begin{figure}
\centering\includegraphics[width=3in]{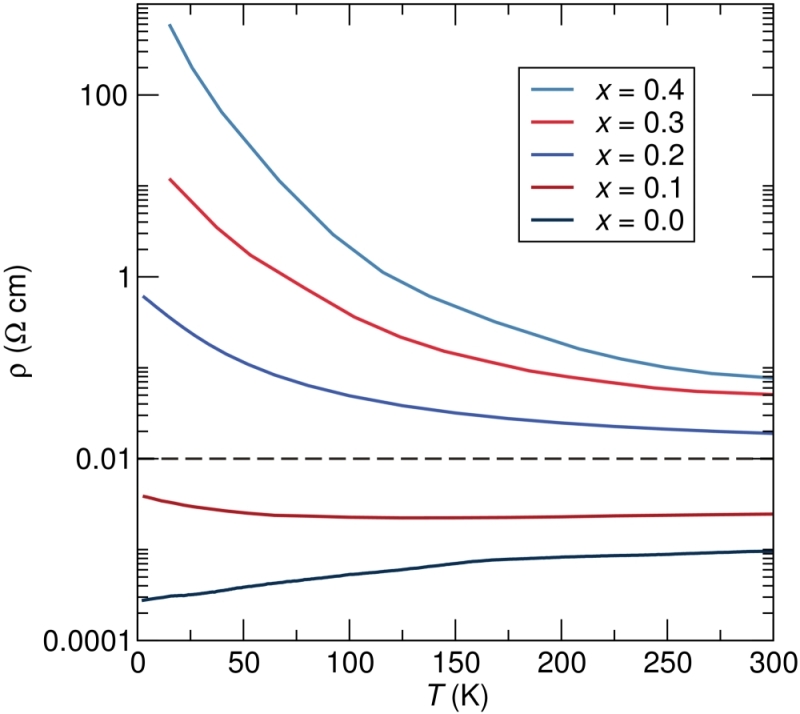}\\
\caption{(Color online) DC electrical resistivity $\rho$ as a function of temperature for the Ru rich samples. A metal-insulator transition occurs as a function of temperature in $x$ = 0.1 while the $x$ = 0.2 is insulating for all temperatures measured.
\label{fig:resistivity}}
\end{figure}

DC electrical resistivity measurements were performed across the solid solution series as presented in FIG.\,\ref{fig:resistivity}. \sro\/ is a metal, exhibiting a positive temperature coefficient of resistivity, and has a resistivity near 10$^{-3}$\,$\Omega$\,cm at 300\,K. Its resistivity displays a kink at the ferromagnetic $T_c$ due to a change in the magnetic scattering of conduction electrons. The $x$ = 0.1 sample has a small upturn in resistivity below 125\,K which is characteristic of weak localization. A compositionally driven metal-to-insulator transition is observed between $x$ = 0.1 and 0.2 where $\rho$ at room temperature becomes greater than the Mott maximum metallic resistivity of $10^{-2}$\,$\Omega$\,cm. For $x$ $\geq$ 0.2 the temperature coefficient of resistivity is always negative.

\subsection{Magnetism}

\begin{table*}
\caption{Magnetic data extracted for \srolro. $T_c$ was determined from peaks in $(\partial\chi/\partial T)$, using the field-cooled $\chi$. $\Theta_{CW}$ and $\mu_{\rm{eff}}$ were obtained by fitting the Curie-Weiss law above 320\,K. $M_{\rm{sat}}$ is the magnetization at 2\,K and 5\,T. \label{tab:mag}}
\centering
\begin{tabular}{lcccccccccccc}
\hline
\hline
$x$ &0.0 &0.1 &0.2 &0.3 &0.4 &0.5 &0.6 &0.8 &0.9\\
\hline
$T_c$ (K)   &160 &100 &50 &24 &16 &9\ \\
$\Theta_{CW}$ (K) &167 &111 &69.4 &47.4 &11.2 &$-$29.4 &$-$42.6 &$-$42.5 &$-$43.2 \\
$\mu_{\rm{eff}}$ ($\mu_B$/Ru) &2.96 &2.77 &2.76 &2.77 &2.82 &3.00 &3.00 &3.02 &3.05 \\
$M_{\rm{sat}}$ ($\mu_B$/Ru) &1.40 &1.13 &0.963 &0.730 &0.358 &0.276 &0.241 &0.251 &0.356 \\
\hline
\hline
\end{tabular}
\end{table*}

Key magnetic data characterizing the solid solution, as determined through analysis of the temperature-dependent magnetic susceptibility, including Curie-Weiss analysis, and measurement of $M$ \textit{vs.} $H$ at 2\,K, and are presented in Table\,\ref{tab:mag}. Curie-Weiss analysis included a temperature independent term for capturing the diagmagnetism of the sample and sample holder background. The value of this parameter was negative and small ($\approx$10$^{-5}$ emu/mol) for all fits, as expected for diamagnetism, though no trend in its value was observed.

\begin{figure}
\centering\includegraphics[width=3in]{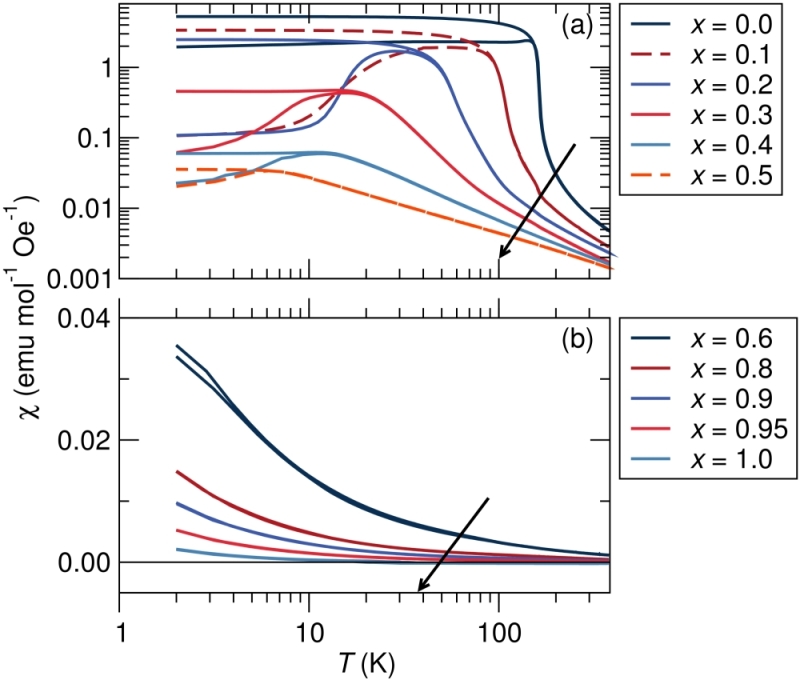}\\
\caption{(Color online) Zero-field cooled and field-cooled magnetic susceptibility collected under a DC field of 1000\,Oe. In (a), samples with $x$ = 0 through $x$ = 0.5 are displayed, and in (b), samples with $x$ = 0.6 through $x$ = 1. \label{fig:chi}}
\end{figure}

\begin{figure}
\centering\includegraphics[width=3in]{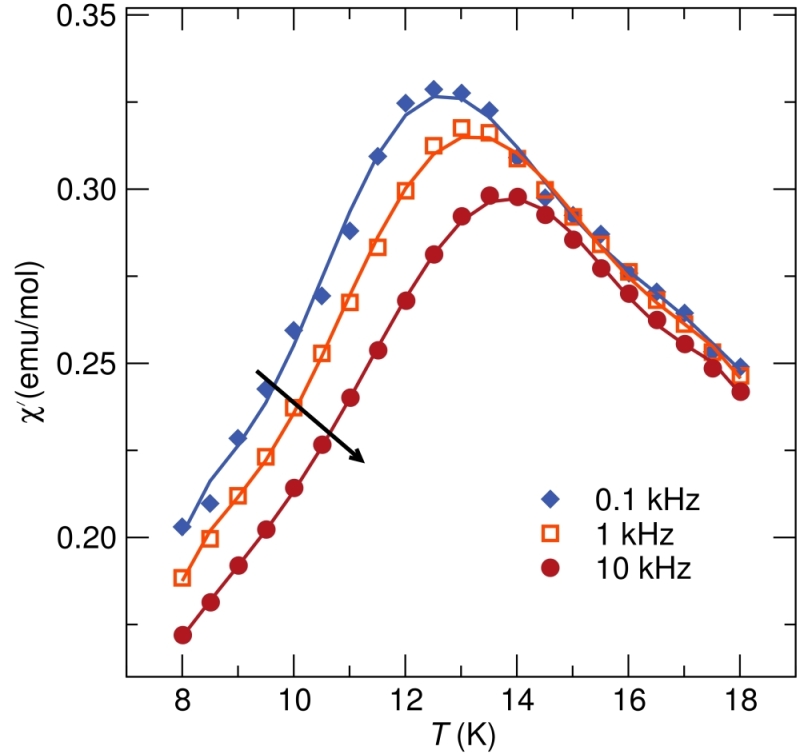}\\
\caption{(Color online) 
AC magnetic susceptibility of the $x$ = 0.3 sample as a function of temperature for different frequencies. Data were collected under zero static field using a 5\,Oe AC field.
\label{fig:acms}}
\end{figure}

Zero-field cooled (ZFC) and field-cooled (FC) magnetic susceptibility data were collected as a function function of temperature under a DC field of 1000\,Oe. The data are displayed in FIG.\,\ref{fig:chi}, with the low-temperature branches in each set of measurements corresponding to the ZFC data. We confirm that the $T_c$ of SrRuO$_3$ is approximately 160\,K as evidenced by the sharp upturn in susceptibility at that temperature. The Curie temperature $T_c$ decreases with \lro\/ substitution and the ordering transition broadens in temperature. For $x >$ 0.5 the system no longer orders and instead has local moment paramagnetic behavior, although the sample with $x$ = 0.6 does display distinct ZFC and FC traces at very low temperatures.  A slight rise in $\chi$ for the $x$ = 0.1 sample occurs at $\approx$160\,K. Although no evidence is seen in the laboratory XRD data, we attribute this to a very small impurity of \sro\/ based on the ordering temperature, and this persists despite continued regrinding, repelletization, and reheating during the preparation procedure. This deviation in susceptibility may be interpreted as Griffiths phase behavior, although we find it more consistent with incomplete reaction and sample inhomogeneity. The system remains ferromagnetic until $x$ = 0.3 where it begins to display the formation of glassy states as observed by the susceptibility reaching a maximum, decreasing, and leveling off to a constant value.  Contrary to previous reports,\cite{Nakamura_JSSC1993,Mary_JSSC1994} the end-member \lro\/ is found to display diamagnetic behavior with negative susceptibility until low temperatures ($\approx$10\,K) at which point paramagnetic impurities or defects become dominant. The absence of diamagnetic behavior in the previous reports was attributed to the presence of small amounts of Rh$^{4+}$ occurring due to impurities or defects. 

To verify the glassy state of the $x$ = 0.3 sample, we conducted frequency-dependent AC magnetic susceptibility measurements, as displayed in FIG.\,\ref{fig:acms}. The dispersion in the peak susceptibility as a function of the frequency of the AC field is characteristic of glassy magnetism.  While not shown, no such frequency dependence of magnetic ordering was seen in the $x$ = 0.1 sample.

\begin{figure}
\centering\includegraphics[width=3in]{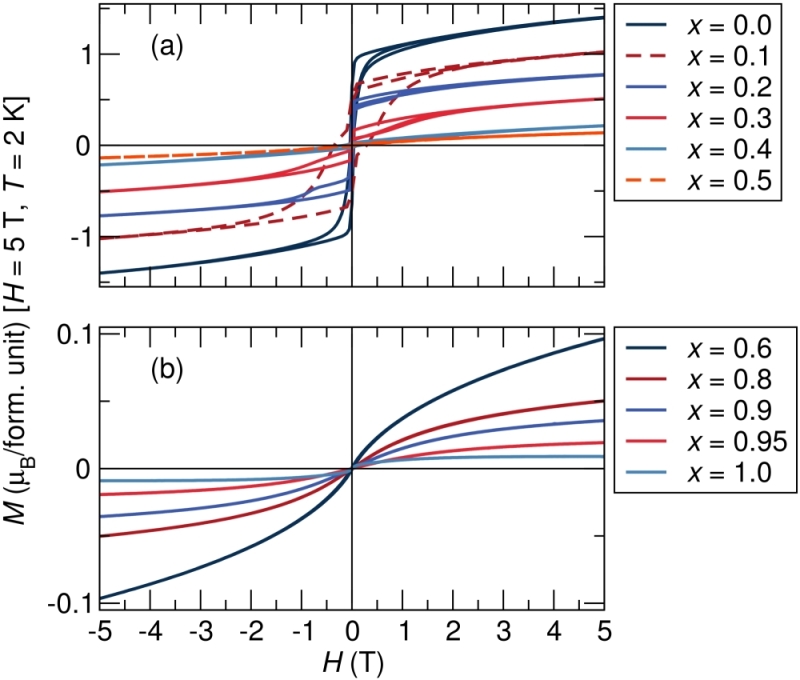}\\
\caption{(Color online) 
Magnetization as a function of applied DC magnetic field at 2\,K. Hysteresis 
is observed for samples that have magnetic order. Data were acquired
in a loop from 0\,T, to 5\,T, to $-$5\,T, and back to 0\,T.
\label{fig:hysteresis}}
\end{figure}

The dependence of magnetization on field for \srolro\/ at 2\,K is shown in FIG.\,\ref{fig:hysteresis}. Hysteresis associated with domain behavior is observed for $x$ $\leq$ 0.5. $M_{\rm{sat}}$ is defined as the magnetization at 2\,K and 5\,T, however none of the samples reach saturation. Sharp discontinuities in magnetization, attributed to powder crystallite rotation under a magnetic field, prevent an analysis of coercivity as a function of substitution.

\begin{figure}
\centering\includegraphics[width=3in]{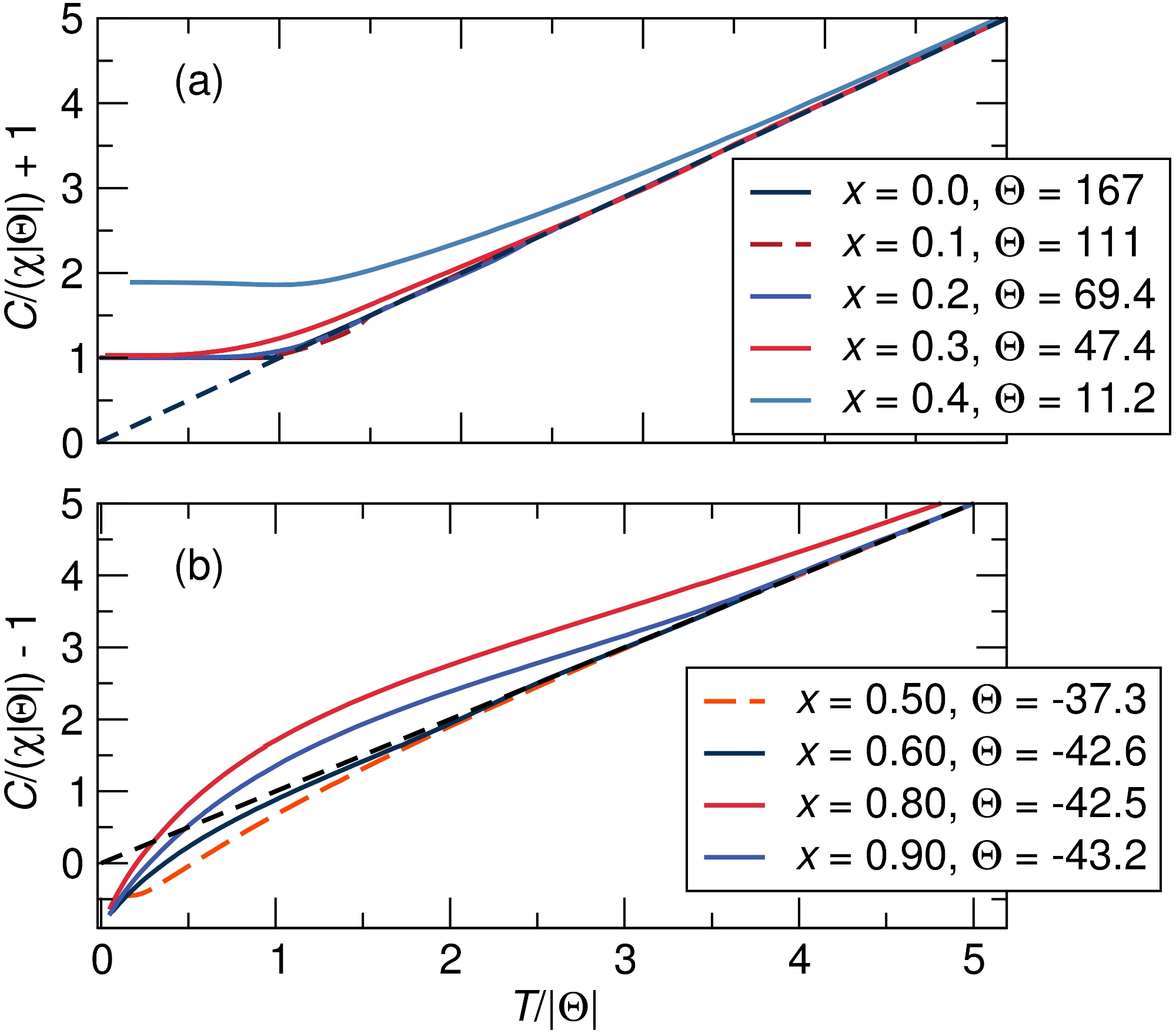}\\
\caption{(Color online) 
Scaled inverse magnetic susceptibility as a function of scaled temperature as described by equation \ref{eqn:CW}. The dashed black line represents ideal Curie-Weiss paramagnetism. The top and bottom panels show data for samples with negative and positive $\Theta_{CW}$ respectively. \label{fig:curie}}
\end{figure}

The Curie-Weiss relation $\chi=C/(T-\Theta_{CW})$ can be recast according to: 

\begin{equation}
\frac{C}{\chi|\Theta_{CW}|}+\mbox{sgn}(\Theta_{CW})=\frac{T}{|\Theta_{CW}|}
\label{eqn:CW}
\end{equation}

\noindent
which allows normalization of susceptibility-temperature plots as shown in FIG.\,\ref{fig:curie}. The utility of such plots has been amply demonstrated in the analysis of other solid-solution systems.\cite{Melot_JPCM2009} For example, the scaled temperature axis allows the frustration index defined $f = \Theta_{CW}/T_c$ to be directly read off. It is seen that all five \sro\/-rich samples in the FIG.\,\ref{fig:curie}(a) order at temperatures corresponding to $T_c \approx \Theta_{CW}$ suggesting that they obey expectations from the Curie-Weiss relationship rather well. At temperatures above the ordering temperature, positive deviations from the ideal Curie-Weiss line reflect the presence of compensated antiferromagnetic short range interactions, while negative deviations reflect uncompensated interactions (ferromagnetism or ferrimagnetism). The ferromagnetic samples ($x$ = 0.0, 0.1, and 0.2) all deviate from ideal Curie-Weiss behavior at their $\Theta_{CW}$ and thus nearly all lie on top of one another on the normalized plot. In contrast, the glassy samples ($x$ = 0.3, 0.4, and 0.5) deviate significantly above their $\Theta_{CW}$. The positive deviation is however difficult to understand, and may have something to do with the nature of the local moments, associated with orbital degeneracy on Ru$^{4+}$. Note that plots for $x \geq 0.4$ may be difficult to interpret due to having $\Theta_{CW}$ near zero. In FIG.\,\ref{fig:curie}(b) samples on the \lro-rich side also display deviations from Curie-Weiss behavior. For larger values of $x$ the majority of interactions become Ru-O-Rh and $\Theta_{CW}$ takes on a small negative value. At low temperatures, the deviations are below the Curie-Weiss line, suggesting that they are not fully compensated, as would be expected for a random alloy.

\begin{figure}
\centering\includegraphics[width=3in]{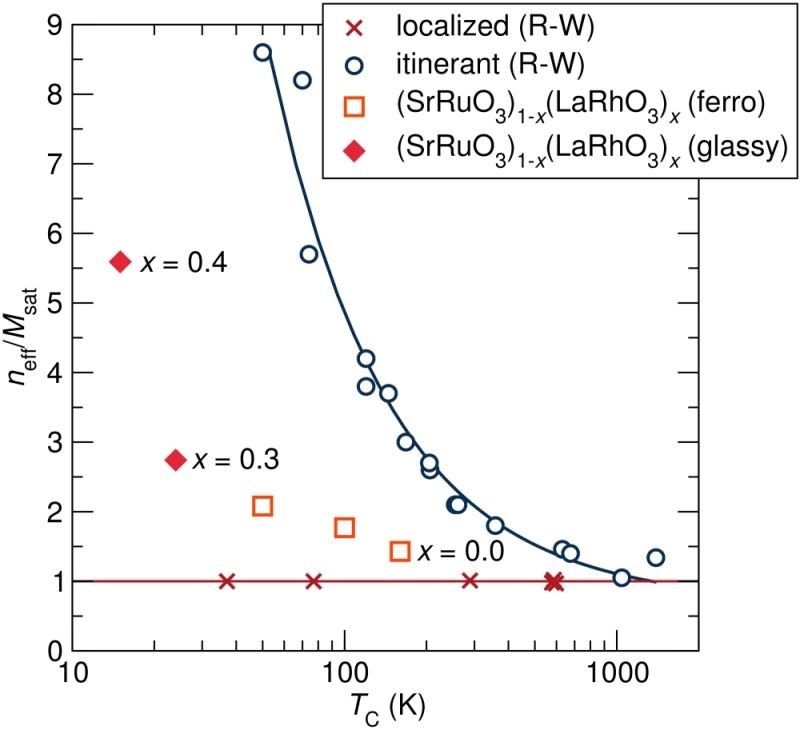}\\
\caption{(Color online) Rhodes-Wohlfarth ratio $n_{\rm{eff}}/M_{\rm{sat}}$ as a function of $T_c$. $n_{\rm{eff}}$/$M_{\rm{sat}}$ = 1, the horizontal red line, indicates localized behavior, with examples from  Rhodes and Wohlfarth\cite{Rhodes_PRSL1963} of CrBr$_3$, MnSb, \textit{etc.}. Ratios lying on the blue curve correspond to itinerant electron ferromagnets, with the curve constructed using data from Rhodes and Wohlfarth\cite{Rhodes_PRSL1963} (ferrous metals, and their alloys with one-another and with Pd). It is seen that \srolro\/ lies between the expectation for local moment and itinerant electron behavior. 
\label{fig:R-W}}
\end{figure}

The fact that the \srolro\/ samples with $x \le 0.4$ follow Curie-Weiss behavior should not by itself be taken as evidence for local-moment behavior.\cite{Takahashi_JPSJ1986} Instead we employ the Rhodes-Wohlfarth ratio, $n_{\rm{eff}}/M_{\rm{sat}}$ to probe the degree of local-moment \textit{versus} itinerant electron behavior in this solid solution series.\cite{Rhodes_PRSL1963} Here $n_{\rm{eff}}$ is the number of unpaired electrons obtained from analysis of the $\mu_{\rm{eff}}$, and for the region of interest, $n_{\rm{eff}}$ is simply $2S$. We find a ratio of 1.43 for \sro\/, using $n_{\rm{eff}}$ and $M_{\rm{sat}}$ determined here, which is larger than previous reports of 1.3 by Fukunaga \textit{et al.}\cite{Fukunaga_JPSJ1994} A ratio of 1.23 is obtained if one uses the spin only $n_{\rm{eff}}$ of 2 for Ru$^{4+}$ and the $M_{\rm{sat}}$ value of 1.63 as found by Bushmeleva \textit{et al.}\cite{Bushmeleva_JMMM2006} from low-temperature neutron diffraction. FIG.\,\ref{fig:R-W} displays the ratio as a function of $T_c$ for \srolro\/ along with data for other well studied systems. We find \srolro\/ to behave intermediate between itinerant and localized for all compositions with magnetic ordering. $n_{\rm{eff}}$/$M_{\rm{sat}}$ deviates further from itinerant behavior and closer to localized behavior with increased \lro\/ substitution. This is expected as there are fewer free carriers as the sample becomes more insulating. It is notable that there is no sharp jump in the ratio at the compositionally driven metal-insulator transition between $x$ = 0.1 and 0.2. The trend with substitution is disrupted as the system turns glassy. 

\begin{figure}
\centering\includegraphics[width=3in]{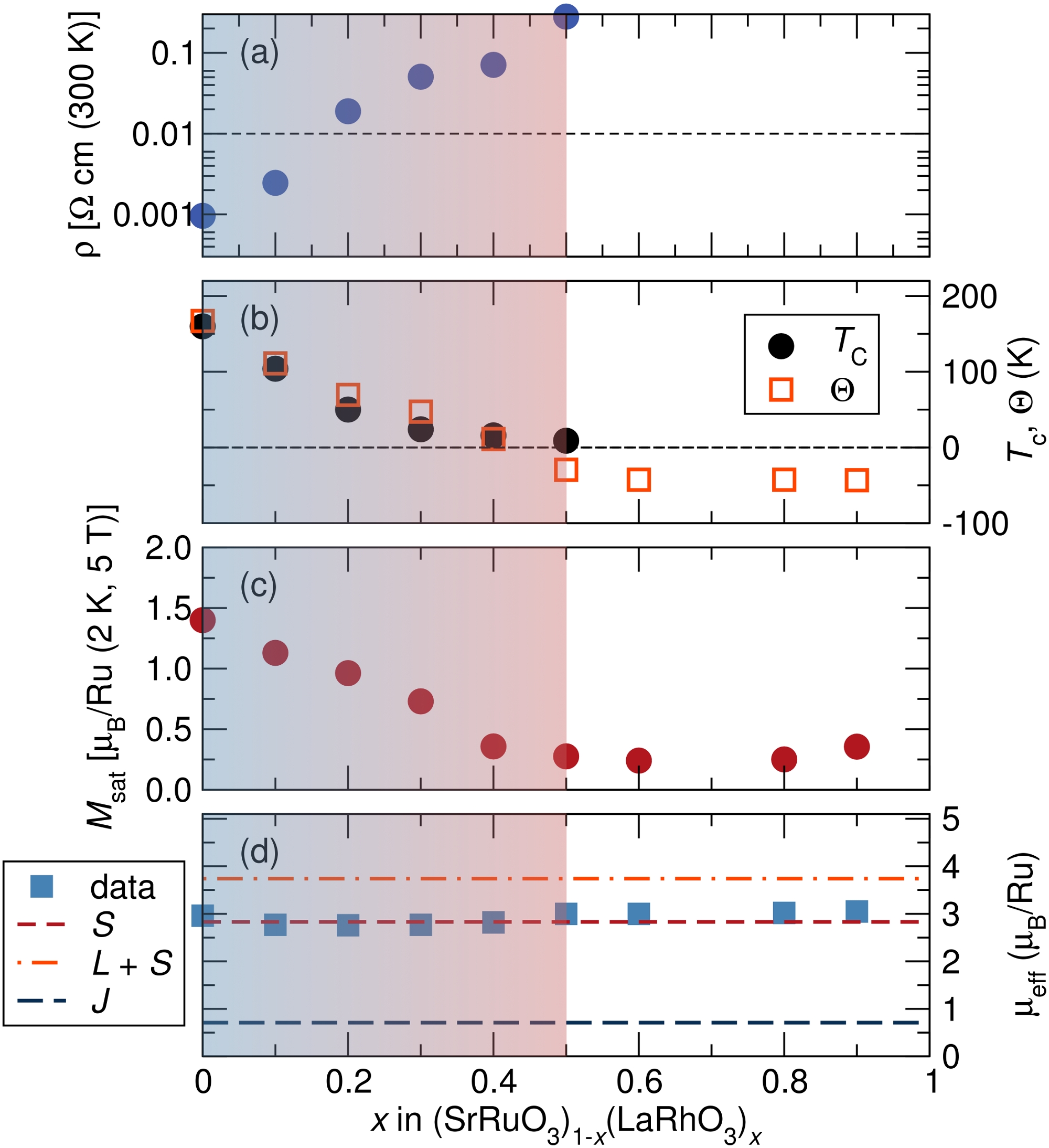}\\
\caption{(Color online) (a) $\rho$(300 K), (b) $T_c$ and $\Theta_{CW}$, (c) $M_{\rm{sat}}$(2\,K, 5\,T), and (d) $\mu_{\rm{eff}}$ as a function of composition. In (a), the black dashed horizontal line at $\rho = 1\times10^{-2}$\,$\Omega$\,cm indicates the Mott maximum metallic resistivity. In (d), the colored dashed horizontal lines are the expected $S$-only, $L+S$, and $J$ values for Ru$^{4+}$. The transition from blue to red indicates the compositionally driven metal-insulator transition, and the entire shaded region corresponds to samples that order magnetically.
\label{fig:phase-dia}}
\end{figure}

FIG.\,\ref{fig:phase-dia} summarizes the findings of this work in the form of a property-composition phase diagram. It displays the electrical resistivity $\rho$ at 300\,K, Curie temperature $T_c$, Curie-Weiss temperature $\Theta_{CW}$, saturation magnetization $M_{\rm{sat}}$ at 2\,K and 5\,T, and effective moment $\mu_{\rm{eff}}$ as a function of \lro\/ substitution, using the data from Table\,\ref{tab:mag}. From panel (a), we see that the formal compositionally-driven metal-to-insulator transition takes place between $x$ = 0.1 and 0.2. Panel (b) shows that $\Theta_{CW}$ first decreases smoothly with \lro\/ substitution as similarly seen for LaFeO$_3$, LaCoO$_3$, and Ca substitution, but then, in contrast to those systems, levels off for large $x$.\cite{Mamchik_PRB2004_A,Mamchik_PRB2004_B,Kanbayasi_JPSJ1978} We have employed $T_c$ values determined from peaks in $(\partial\chi/\partial T)$ using field-cooled $\chi$. For small $0\le x < 0.5$, $T_c \approx \Theta_{CW}$ suggesting that Curie-Weiss behavior is obeyed. Around $x$ = 0.5, the value of $\Theta_{CW}$ becomes small and negative, and this is the region beyond which there is no evidence for magnetic ordering. $T_c$ decreases with \lro\/ substitution, as might be expected given that magnetic Ru$^{4+}$ are diluted with increasing $x$, but the rate at which the decrease takes place is only slightly faster that what is seen in the \srocro\/ system.\cite{He_PRB2000} The Ru--O--Ru bond angles also decrease with $x$ in a similar manner as is seen in \srocro\/ and the suggestion is that structural effects as $x$ increases are perhaps as important as the effect of filling $t_{2g}$. Comparisons can be made with the Co$_{1-x}$Fe$_x$S$_2$ pyrite system which exhibits Stoner ferromagnetism.\cite{Jarrett_PRL1968,Ramesha_PRB2004,Guo_PRB2010} There, it is observed that $T_c$ remains constant over a large range of $x$ while in \srolro\ we find that $T_c$ drops rapidly with \lro\ substitution. This may suggest that \srolro\/ does not strictly follow Stoner-Wohlfarth band ferromagnetism. Further, the $x$ = 0.2 sample is a good ferromagnet despite being an electrical insulator. $M_{\rm{sat}}$ \textit{per} Ru decreases with \lro\/ substitution as seen in FIG.\,\ref{fig:phase-dia}(c), in the region of magnetic ordering, and then is more-or-less flat with $x$. As previously reported, \sro\/ has a $\mu_{\rm{eff}}$ equal to the spin only $S$ value for Ru$^{4+}$ as the orbital contribution expected for an octahedral $d^4$ cation is quenched out.\cite{Randall_JACS1959} As shown in FIG.\,\ref{fig:phase-dia}(d), this does not change with \lro\/ substitution, as $\mu_{\rm{eff}}$ remains essentially constant across the solid solution as the Ru coordination environment remains unchanged.

\section{Conclusions}

This study of the electrical and magnetic properties of \srolro\/ has enabled some important observations and conclusions regarding the magnetism in \sro\/ to be made.  We note the magnetism in the solid solution does not require metallic conduction in order to persist. While the Rhodes-Wohlfarth ratio has previously been reported for \sro\/ and provided as evidence of intermediate behavior between localized and itinerant, we demonstrate that alloying with a diamagnetic semiconductor pushes the behavior to become more localized. Comparison with \srocro\/ reveals that octahedral tilting is nearly as effective at disrupting ferromagnetism in \srolro\/ as is the dilution of magnetism achieved by filling $t_{2g}$.

\section{Acknowledgments}
We gratefully acknowledge useful discussions with James R. Neilson and Brent C. Melot. This project is supported through a Materials World Network Award from the NSF (DMR 0909180) in Santa Barbara, and from the EPSRC (EP/G065314/1) in Liverpool. We acknowledge the use of MRL Central Facilities which are supported by the MRSEC Program of the NSF under Award No. DMR05-20415; a member of the NSF-funded Materials Research Facilities Network (www.mrfn.org). Use of data from the 11-BM beamline at the Advanced Photon Source was supported by the U.S. Department of Energy, Office of Science, Office of Basic Energy Sciences, under Contract No. DE-AC02-06CH11357.

\clearpage

\bibliography{PhillipBarton}

\clearpage

\end{document}